\documentclass{svjour2}

\usepackage{mathptmx}
\usepackage{amsmath}
\usepackage{amssymb}
\usepackage{dsfont}
\usepackage{relsize}
\usepackage{nicefrac}
\usepackage{psfrag}
\usepackage{graphicx}

\newcommand{\nn}{\nonumber\\}
\newcommand{\set}[1]{{\{#1\}}}
\newcommand{\Sp}{\mbox{Tr}\,}
\newcommand{\e}[1]{\mbox{e}^{#1}}
\newcommand{\de}{\mbox{d}}
\newcommand{\LL}{\mathds{L}}
\newcommand{\maxm}{\max}
\newcommand{\bra}[1]{\langle{#1}|}
\newcommand{\ket}[1]{|{#1}\rangle}
\newcommand{\emb}[1]{\mbox{e}^{\mbox{-}\beta #1}}
\newcommand{\hX}{\vec{X}}
\newcommand{\hY}{\vec{Y}}
\newcommand{\hZ}{\vec{Z}}
\newcommand{\hH}{\vec{H}}
\newcommand{\hT}{\vec{T}}
\newcommand{\hu}{\vec{1}}
\newcommand{\sz}{\vec{s}^z}
\newcommand{\sx}{\vec{s}^x}
\newcommand{\sy}{\vec{s}^y}
\newcommand{\mot}[2]{\overset{#2}{\underset{#1}{\bigotimes}}}
\newcommand{\oty}{{\mathcal O}(t_y^2)}
\newcommand{\overdelta}{\delta}
\newcommand{\smallhalf}{\frac{\mbox{\tiny 1}}{\mbox{\tiny 2}}}
\newcommand{\nneqjjpo}[2]{{
\def\arraystretch{0.0}
\setlength\arraycolsep{0.0pt}
\!\begin{array}{ll}
\scriptscriptstyle #1\neq #2\\{\scriptscriptstyle #1\neq 
#2+1}\\\phantom{n}\\\phantom{n}\\\phantom{n}
\end{array}
}}
\newcommand{\sidesum}[1]{
{\vphantom{\sum}}^{#1}}

\begin{document}

\title{Strong-interaction approximation for transfer matrix method}

\author{Oles Zaburannyi}

\institute{
O. Zaburannyi \at 
Institute for Condensed Matter Physics, 
National Academy of Sciences of Ukraine, 
1 Svientsitskii Street, L'viv-11, 79011, Ukraine\\
\email{zab@icmp.lviv.ua}
}

\date{\today}

\maketitle

\begin{abstract}
Using transfer-matrix method a correspondence between $2D$
classical spin systems ($2D$ Ising model and six-vertex model) and
$1D$ quantum spin systems is considered.
We find the transfer matrix in two limits - 
in a well-known strong-anisotropy limit and
a novel strong-interaction limit.
In contrast to the usual strong-anisotropy approximation,
within the strong-interaction approximation
we take into account the non-commutativity of transfer-matrix components.
The latter approximation is valid for low temperatures
or strong interaction in one spatial dimension. 
We observe that the Hamiltonian
of the corresponding quantum chains contains
multispin interactions.
\keywords{transfer matrix \and quantum spin chains}
\PACS{
05.50.+q \and% Lattice theory and statistics
05.30.Rt \and% Quantum phase transitions
75.10.Pq% Spin chain models
}
\end{abstract}
\section{Introduction}
\label{sec:Introduction}
Transfer matrix method allows to find thermodynamic properties 
for many low-dimensional models \cite{Baxter_1982}, \cite{Lieb_1967:prl19-108}
and shows thermodynamic equivalence for many of them 
\cite{Rommelse_Nijs_1987:prl59-2578},
\cite{Carlon_Mazzeo_vanBeijeren_1997:prb55-757}.
Important feature of the method is the possibility to establish
a relation between thermodynamic of $d$-dimensional classical models and
ground state of $(d-1)$-dimensional quantum models \cite{Kogut_1979:rmp51-659}
under the assumption of strongly anisotropic interactions in a classical system.

In the present paper we discuss the conditions which are imposed on a classical
system within strong-anisotropy approximation.
Besides that, a new strong-interaction approximation
which requires weaker conditions is introduced.
For the new approximation the transfer matrix is written in the symmetric form
$T=\e{t_x X}\e{t_y Y}\e{t_x X}$, see \eqref{eqn:Tnoncommutative},
and the expansion in nested commutators similar 
to the Baker-Campbell-Hausdorff formula is performed.
The quantum Hamiltonian can be found explicitly only if the nested commutator
of an arbitrary order $[X,\ldots[X,Y]\ldots]$ can be written in a
specific form, (see  \ref{eqn:Z_commutatorexplicitly}).
The strong-interaction limit reduces to the strong-anisotropy limit if we
assume $[X,Y]=0$. The strong-anisotropy limit also can be obtained
by applying the exponential operator decomposition
technique \cite{Suzuki_1990:pla146-319},
\cite{Kobayashi_Hatano_Suzuki_1994:physa211-234}.
The main difference between the operator decomposition technique and the 
strong-interaction limit is that in the latter case we must calculate
nested commutators instead of assuming that $t_x \propto t_y$.
For the $2D$ spin-$\smallhalf$ classical Ising model the strong-interaction
limit leads to the appearance of three-spin interactions in the resulting
quantum chain. For the six-vertex model the new approximation 
gives the $XXZ$ chain with four-spin interactions.

We will use the transfer matrix method for classical $2D$
models \cite{Lieb_Niemeijer_Vertogen_1970} with 
the total energy that can be represented as a sum over rows
\begin{align}
\label{eqn:classical}
E=\sum_{m=1}^M E(\xi_m,\xi_{m+1}),
\end{align}
where $\xi_m$ is a variable defined on the row $m$ with $\LL$ possible 
values. The free energy of the classical $2D$ model (per row) can be written
in the form
\begin{align}
\label{eqn:TpowerM}
f = - \lim_{M\rightarrow \infty} \frac{1}{\beta M} \ln \Sp T^M =
- \tfrac{1}{\beta} \ln \lambda_{\maxm},
\end{align}
where $T$ is the transfer matrix with elements 
$T_{\xi_m,\xi_{m+1}}=\emb{E(\xi_m,\xi_{m+1})}$ and the maximal eigenvalue 
$\lambda_{\maxm}$ which is real, unique and positive
according to the Perron-Frobenius theorem \cite{Berman_Plemmons_1994} for 
the positive symmetric matrix $T$.
We can consider $\LL$ configurations as orthonormal basis
$\{\ket{\xi}\}_{\xi=\overline{1,\LL}}$ of $\LL$-dimensional Hilbert space.
Each configuration $\xi$ corresponds to a base-vector in the Hilbert space.
The transfer matrix $T$ corresponds to some operator
$\hT = \sum_{\xi,\xi'=1}^\LL \ket{\xi} T_{\xi,\xi'} \bra{\xi'}$.
Let us introduce the Hamiltonian of a quantum system, which is defined
by the logarithm of the operator $\hT$
\begin{align}
\label{eqn:H}
\vec{H} = - \tfrac{1}{\beta} \ln \hT.
\end{align}
The ground state energy for the quantum system described by 
Hamiltonian \eqref{eqn:H} is equal to the free energy \eqref{eqn:TpowerM},
\begin{align}
\label{eqn:e0}
e_0 = \lim_{\beta_q\to \infty} 
\frac{\Sp \left( 
\e{\mbox{-}\beta_q \vec{H}} \vec{H}
\right) }{\Sp \e{\mbox{-}\beta_q \vec{H}} }
= - \tfrac{1}{\beta} \lim_{\beta_q \to \infty} 
\frac{\Sp \left( \hT^{\frac{\beta_q}{\beta}} \ln \hT \right) }
{\Sp \hT^\frac{\beta_q}{\beta} }
= - \tfrac{1}{\beta} \ln \lambda_{\maxm} = f.
\end{align}
Here $\beta_q$ is the inverse temperature of quantum system \eqref{eqn:H} and
it must be distinguished from the inverse temperature of classical
system \eqref{eqn:classical} $\beta$. We will elaborate approximate method
which allow to find Hamiltonian \eqref{eqn:H} for $2D$ classical 
systems \eqref{eqn:classical} at low temperatures or with strong interaction 
in one spatial dimension.

The paper is organized as following. In the next (second) section
the strong-anisotropy and the strong-interaction approximations in general
for a $2D$ classical system are presented.
In the third and fourth sections the methods are
applied for the $2D$ Ising model and the six-vertex model respectively.
Quantum Hamiltonians are obtained there.
Finally, in section five we compare the approximations
by calculating critical temperatures for classical models.
Important calculations which we use for the strong-interaction approximation 
are collected in Appendix.
\section{Approximations for transfer matrix}
\label{sec:Approximations}
In general, for an arbitrary $2D$ classical system with the total energy 
\eqref{eqn:classical} it is impossible to find the transfer matrix logarithm 
and write down the Hamiltonian \eqref{eqn:H} explicitly.
In this section we consider two approximate methods used to find the quantum 
system Hamiltonian \eqref{eqn:H}.
Both of them are based on division of energy $E$ \eqref{eqn:classical} 
into two parts, $E_x$ which depends on configuration $\xi$ 
of single row only and $E_y$ which depends on two configurations 
$\xi$ and $\xi'$ of successive rows.
There are many possible divisions, but we choose the 
division symmetric with respect to two neighboring rows
\begin{align}
\label{eqn:exey}
E(\xi,\xi') = \frac{E_x(\xi)}{2} + E_y(\xi,\xi') + \frac{E_x(\xi')}{2}.
\end{align}
Here $E_x$ is the energy of a single row with configuration 
$\xi$ and $E_y(\xi,\xi')$ is the energy of the interaction between
two rows with configurations $\xi$ and $\xi'$.
Division \eqref{eqn:exey} allows us to write
$T_{\xi,\xi'} = T_\xi^x T_{\xi,\xi'}^y T_{\xi'}^x$, where
$T_\xi^x = \emb{\frac{E_x(\xi)}{2}}$, $T_{\xi,\xi'}^y=\emb{E_y(\xi,\xi')}$.
Accordingly we can rewrite $T$ as a matrix product, 
$T=T^x T^y T^x$, and $\hT$ as an operator product,
\begin{align}
\hT = \sum_{\xi,\xi'=1}^\LL \ket{\xi} T_{\xi,\xi'} \bra{\xi'}
= \hT^x \hT^y \hT^x,
\end{align}
where $\hT^x = \sum_{\xi=1}^\LL \ket{\xi} T_{\xi}^x \bra{\xi}$ is the
diagonal operator and
$\hT^y=\sum_{\xi,\xi'=1}^\LL \ket{\xi} T_{\xi,\xi'}^y \bra{\xi'}$ 
is the off-diagonal operator.

For further progress we have to rewrite operator $\hT^y$ in a slightly 
different form. It is convenient to assume
$E_y(\xi,\xi) = 0$ (that can be always 
achieved by including nonzero value $E_y(\xi,\xi)$ into $E_x(\xi)$ or by
shifting all energies by a constant value).
That makes diagonal elements 
of $T^y$ equal to unity, $T^y_{\xi,\xi}=1$. Among all off-diagonal
elements of $\hT^y$ we distinguish the elements which are proportional to
some parameter $t_y$ (which we will demand later to be small) 
and collect these elements in operator $\hY$
\begin{align}
\hY_{\xi,\xi'}=
\left\{
\begin{array}{ll}
T^y_{\xi,\xi'},&\mbox{ if } T^y_{\xi,\xi'}\propto t_y\\
0,&\mbox{ in other cases }.
\end{array}
\right.
\end{align}
All another off-diagonal elements which are $\oty$ we collect in operator 
$\hY'$.
Summing up
\begin{align}
\label{eqn:TXY}
\hT = \e{t_x \hX} \left( \vec{1} + t_y \hY + \oty\hY^\prime \right) 
\e{t_x \hX},
\end{align}
where diagonal part $\hT^x$ is rewritten in the form $\hT^x = \e{t_x \hX}$
Until now no approximation has been made. We only assume that 
the parameters $t_x$ and $t_y$ exist. For both approaches, i.e.,
strong-anisotropy limit and strong-interaction limit
it will be later required that $t_y$ is small and we will neglect the
terms $\oty$.
For each specific model the individual operator $\hY$ should be constructed
bearing this requirement in mind.
\vspace{1em}\\
{\bf \thesection.1 The strong-anisotropy limit} can be 
introduced by neglecting the terms $\oty$ and by a naive assumption - 
commutativity of $\hX$ and $\hY$ in the expression for $\hT$:
\begin{align}
\label{eqn:Tnoncommutative}
\hT =
\e{t_x \hX} \left( \vec{1} + t_y \hY + \oty\hY^\prime  \right) \e{t_x \hX}
=
\e{t_x \hX} \e{t_y \hY } \e{t_x \hX} + \oty
\underset{\mathsmaller{[\hX,\hY]=0}}{\approx}
\e{2t_x \hX + t_y \hY }.
\end{align}
More precisely, the last approximation in \eqref{eqn:Tnoncommutative} can be
obtained by series expansions of the exponents and neglecting 
the terms ${\mathcal O}(t_x^2)$ and $\oty$,
\begin{align}
\label{eqn:Tnoncommutativecondition}
\hT =
\e{t_x \hX} \left( \vec{1} + t_y \hY + \oty\hY^\prime  \right) \e{t_x \hX}
=
\e{2t_x \hX + t_y \hY } + {\mathcal O}(t_x^2) + \oty.
\end{align}
Finally, the Hamiltonian in the strong-anisotropy limit has the form
\begin{align}
\label{eqn:Hsal}
\vec{H}_{sal} = - \tfrac{1}{\beta} \ln \hT = 
- \frac{2t_x}{\beta} \hX - \frac{t_y}{\beta} \hY.
\end{align}

For the strong-anisotropy approximation it is often assumed that
$t_x \propto t_y$ and $\oty=0$. This conditions impose some relations on
the classical system parameters \eqref{eqn:classical}.
Assumption $t_x \propto t_y$ is good to explain why approximation is 
called \textit{strong-anisotropy} limit.
If we consider the simplest case for model \eqref{eqn:classical} 
with all in-row energies proportional to $e_x$, $E_x(\xi)\propto e_x$,
and all inter-row interaction energies for different configurations not smaller
than $e_y$, $E_y(\xi,\xi')\propto e_y,\;\;\xi\neq\xi'$, then the
condition $t_x \propto t_y$ reads
\begin{align}
\label{eqn:anisotropy}
\beta e_x \propto  \emb{e_y}.
\end{align}
By demanding $\emb{e_y}$ to be small we suppose $e_y$ to be large and
$e_x$ to be small. It should be noted that by demanding $\emb{e_y}$ to be small 
we also assume that $e_y>0$. Conditions $t_x\propto t_y$, $\oty=0$ is 
weakest from a set of assumptions $t_x\in{\mathcal O}(t_y)$, $\oty=0$ 
which allow to apply strong-anisotropy approach.
\vspace{1em}\\
{\bf \thesection.2 The strong-interaction limit},
in contrast to the strong-anisotropy limit,
demands only $t_y$ to be small and does not put any restrictions on $t_x$.
We can rewrite the expression for transfer matrix \eqref{eqn:TXY} in the form
\begin{align}
\label{eqn:Zintroducing}
\hT = \e{t_x \hX} (1+  t_y \hY + \oty) \e{t_x \hX} 
= \e{2t_x \hX+t_y (\hY+\hZ)} + \oty
\end{align}
where $\hZ$ is an unknown operator
defined by the function $\hZ=Z(t_x\hX,\hY)$. The function Z can be expand
as series in the nested commutators of operators $\hX$ and $\hY$ (see Appendix).
If a general nested commutator $[\hX,\hY]_n=[\hX,[\hX,\hY]_{n-1}]$
can be presented explicitly via some
operators $\vec{L}$, $\vec{z}$, $\vec{R}$ 
(see \ref{eqn:Z_commutatorexplicitly}),
all calculations can be performed to the very end
and $\hZ$ can be found explicitly.
Thus, the Hamiltonian in the strong-interaction limit takes the form
\begin{align}
\label{eqn:Hsil}
\hH_{sil} =  - \frac{2 t_x}{\beta}  \hX - \frac{t_y}{\beta} \hY - 
\frac{t_y}{\beta} \sum_{q=1}^Q  \sum_{l=0}^{P-1} \vec{L}_{l,q}
\mathcal{A}_{P,l}(t_x \vec{z}_{l,q}) \vec{R}_{l,q},
\end{align}
where $\mathcal{A}_{P,l}(x)
=
\frac{1}{P} \sum_{p=0}^{P-1}
\e{\mbox{-}\frac{2i\pi p l}{P}}
\mathcal{A}\left(x \e{\frac{i \pi p}{P}}\right)$,
$\mathcal{A}(x)=\frac{x}{\sinh x}-1$.
The strong-anisotropy limit \eqref{eqn:Hsal} follows from the strong-interaction
limit \eqref{eqn:Hsil} as expected. If we assume $t_x \propto t_y$ and neglect
the terms $\mathcal{O}(t_y^2)$ the last term in \eqref{eqn:Hsil} 
vanishes since $\mathcal{A}_{P,l}(x) \in \mathcal{O}(x^2)$.
\section{Two-dimensional Ising model}
\label{sec:Ising}
In this section both approximations will be used for the
$2D$ classical Ising model in order to find the corresponding 
$1D$ quantum system.
The classical $2D$ Ising model is described by the Hamiltonian
\begin{align}
\label{eqn:2DIsing}
E= - \sum_{m=1}^M \sum_{n=1}^N J_x \sigma_{m,n} \sigma_{m,n+1} + J_y 
\sigma_{m,n} \sigma_{m+1,n},
\end{align}
where $\sigma_{m,n}$ assumes two values $\pm \tfrac{1}{2}$.
A configuration in each row $m$ is defined by a set of
variables $\{\sigma_{m,n}\}_{n=\overline{1,N}}$ and it takes $\LL = 2^N$ 
possible values.
We will denote the configurations
on two successive rows
by $\{\sigma_n\}$ and $\{\sigma'_n\}$
and use the following notation
\begin{align}
E_x(\{\sigma_n\})= 
- \sum_{n=1}^N J_x \sigma_{n} \sigma_{n+1},\nn
E_y(\{\sigma_n\},\{\sigma'_n\}) = 
- \sum_{n=1}^N J_y \sigma_{n} \sigma'_{n}.
\end{align}
The Hilbert space can be spanned by the basis
$\{\ket{\xi}\}_{\xi=\overline{1,2^N}}
=\{\otimes_{n=1}^N \ket{\sigma_n}\}_{\sigma_1\ldots\sigma_N=\pm\frac{1}{2}}$.
We can establish a correspondence between the states $\ket{\sigma}$, 
$\sigma = \pm \frac{1}{2}$ and the eigenvectors of the spin operator
$\sz = \sum_{\sigma} \ket{\sigma}\sigma\bra{\sigma}$.
Here and further on
$\sum_{\sigma}$ denotes $\sum_{\sigma=\pm \frac{1}{2}}$.
The diagonal part of the transfer matrix $\hX$ has the form
\begin{align}
\label{eqn:isingtx}
\hT^x =
\sum\limits_{\sigma_1}
\ldots
\sum\limits_{\sigma_N}
\mot{n=1}{N} \ket{\sigma_n}
\e{\beta\frac{J_x}{2} \sum\limits_{j=1}^N \sigma_j \sigma_{j+1}}
\mot{n=1}{N} \bra{\sigma_n}
\nn
=
\e{\beta\frac{J_x}{2} \sum\limits_{j=1}^N
\sum\limits_{\sigma_1}
\ldots
\sum\limits_{\sigma_N}
\mot{n=1}{N} \ket{\sigma_n} \sigma_j \sigma_{j+1} \mot{n=1}{N} \bra{\sigma_n}}
=
\e{\beta\frac{J_x}{2} \sum\limits_{j=1}^N \sz_j \sz_{j+1}},
\end{align}
where we omitted the direct products of the identity
operators. The off-diagonal operator $\hT^y$ takes the form
\begin{align}
\label{eqn:isingty}
\hT^y=
\sum\limits_{\sigma_1}
\ldots
\sum\limits_{\sigma_N}
\sum\limits_{\sigma^\prime_1}
\ldots
\sum\limits_{\sigma^\prime_N}
\mot{n=1}{N} \ket{\sigma_n}
\e{\beta J_y \sum_{j=1}^N \sigma_j \sigma'_j}
\mot{n=1}{N} \bra{\sigma'_n}.
\end{align}
In contrast to the case of diagonal operator $\hT^x$, 
Eq. \eqref{eqn:isingtx}, we can not bring the
direct products under the exponent. To construct the operator $\hY$, we 
classify all matrix elements $\hT^y_{\{\sigma_n\},\{\sigma'_n\}}$ into 
three groups depending upon a number of
different variables in $\{\sigma_n\}$ and $\{\sigma'_n\}$.
In the first group we collect all elements for which the rows 
$\{\sigma_n\},\{\sigma'_n\}$ are the same, i.e. the diagonal elements 
of $\hT^y$. Diagonal elements are equal to $\e{\frac{N\beta J_y}{4}}$, as it
was discussed above, we can obtain 
$\hT^y_{\{\sigma_n\},\{\sigma_n\}}=1$ after shifting 
the energies by an appropriate quantity. 
All the matrix elements where the variables only on one
cite differ, i.e., 
$\left\{\sigma_n=\sigma'_n\right\}_{n=\overline{1,N},n\neq j},
\sigma_j=-\sigma^\prime_j$ will form the second group.
In this case the interaction energy increases by 
$\frac{J_y}{2}$ and the matrix element of $\hT^y$ 
after energy shifting is equal to $\emb{\frac{J_y}{2}}$.
The third group consists of all other matrix elements
for the states $\{\sigma_n\}$, $\{\sigma'_n\}$  
which are different on two or more sites.
These elements are equal to $\left(\emb{\frac{J_y}{2}}\right)^r$,
where $r$ is the number of sites with different variables 
$\sigma_n$, $\sigma'_n$.
Now, it can be seen how the small parameter $t_y$ should be set.
If $t_y=\emb{\frac{J_y}{2}}$ is small, 
we can construct an operator $\hY$ which 
has all matrix elements equal to zero except the elements 
between the states which differ by one variable $\sigma_j$ (second group).
All other matrix elements (third group) which are of order $\oty$ 
may be included into a non-important operator $\hY^\prime$.
Formally it can be done by rewriting the sum \eqref{eqn:isingty}
as follows:
\begin{align}
\label{eqn:isingty2}
\hT^y=&
\sum\limits_{\sigma_1}
\ldots
\sum\limits_{\sigma_N}
\mot{n=1}{N}\ket{\sigma_n}
\mot{n=1}{N}\bra{\sigma_n}
\nn
+&
t_y
\sum_{j=1}^N
\sum\limits_{\sigma_1}
\ldots
\sum\limits_{\sigma_N}
\mot{n=1}{j-1}\ket{\sigma_n}
\otimes \ket{\sigma_j}
\mot{n=j+1}{N}\ket{\sigma_n}
\mot{n=1}{j-1}\bra{\sigma_n}
\otimes \bra{-\sigma_j}
\mot{n=j+1}{N}\bra{\sigma_n}
\nn
+&
\sum_{\substack{\textrm{\tiny all other}\\\textrm{\tiny configurations}}}
\emb{E_y(\set{\sigma}, \set{\sigma'})}
\mot{n=1}{N}\ket{\sigma_n}
\mot{n=1}{N}\bra{\sigma'_n}
\nn
=&\hu + t_y \hY + \oty\hY^\prime.
\end{align}
(We notice that in order to calculate higher approximations with respect to 
$t_y$ operator $\hY^\prime$ should be presented as series with respect 
to $t_y^r$, $r\ge2$. The term proportional to $t_y^r$ contains
$\binom{N}{r}$ matrix elements of $\hT^y$).
The operator $\hY$ can be 
easily identified in terms of spin operators,
\begin{align}
\label{eqn:isingty3}
\hY&=
\sum_{j=1}^N
\sum\limits_{\sigma_1}
\ldots
\sum\limits_{\sigma_N}
\mot{n=1}{j-1}\ket{\sigma_n}
\otimes \ket{\sigma_j}
\mot{n=j+1}{N}\ket{\sigma_n}
\mot{n=1}{j-1}\bra{\sigma_n}
\otimes \bra{-\sigma_j}
\mot{n=j+1}{N}\bra{\sigma_n}
\nn
&=2 \sum_{j=1}^N \sx_j,
\end{align}
where 
$\sx = \frac{1}{2}\sum_{\sigma} \ket{\sigma}\bra{-\sigma}$.
From Eqs. \eqref{eqn:isingtx} and \eqref{eqn:isingty3} we see that
$t_x$, $t_y$, $\hX$, $\hY$
can be written as
\begin{align}
\label{eqn:isingttxtyhxhy}
t_x=\frac{\beta J_x}{2},\;\;
t_y=\emb{\frac{J_y}{2}},\;\;
\hX=\sum\limits_{j=1}^N \sz_j \sz_{j+1},\;\;
\hY=\sum\limits_{j=1}^N 2 \sx_j.
\end{align}
In the strong-anisotropy limit we obtain the Hamiltonian of the quantum Ising 
chain in transverse field
\begin{align}
\label{eqn:isingHsal}
\hH_{sal} = -\frac{2}{\beta} t_x \hX - \frac{t_y}{\beta} \hY =
- \sum_{j=1}^N J_x \sz_j \sz_{j+1} + \frac{2}{\beta}\emb{\frac{J_y}{2}} \sx_j.
\end{align}

To obtain the Hamiltonian in the strong-interaction limit (i.e., $t_y$ is small) 
we have to consider the commutators $[\hX,\hY]_{2k}$
\begin{align}
\label{eqn:isingcommutators}
[\hX,\hY]_{2k} = 2\sum_{j=1}^N \sx_j 
\left(\frac{1}{2}+2\sz_{j-1}\sz_{j+1}\right),\;\;k\geq 1.
\end{align}
Comparing Eqs. \eqref{eqn:isingcommutators} and \eqref{eqn:Z_commutatorexplicitly} 
we find that we have $P,Q=1$, $l=0$, 
$\vec{L}_{0,1}\!=\!\sum\limits_{j}\sx_j(1+4\sz_{j-1}\sz_{j+1})$,
$\vec{z}_{0,1}=1$,
$\vec{R}_{0,1}=1$ and only one function $\mathcal{A}_{1,0}(t_x)=\mathcal{A}(t_x)$ 
in \eqref{eqn:apl} have to be calculated.
Therefore, from \eqref{eqn:Hsil}
\begin{align}
\label{eqn:isingHsil}
\hH_{sil} = - \sum_{j=1}^N J_x \sz_j \sz_{j+1} +
\frac{2}{\beta}\emb{\frac{J_y}{2}} \sx_j \left(1 +
 \tfrac{1}{2}\mathcal{A}\left(\tfrac{\beta J_x}{2}\right)
+2\mathcal{A}\left(\tfrac{\beta J_x}{2}\right)\sz_{j-1}\sz_{j+1}\right).
\end{align}
In comparison to \eqref{eqn:isingHsal} the strong-interaction 
approximation \eqref{eqn:isingHsil} implies renormalized transverse field
$\left[1\rightarrow 
1+\tfrac{1}{2}\mathcal{A}\left(\tfrac{\beta J_x}{2}\right)\right]$
and additional three-spin interactions of 
$\sz_{j-1}\sx_j\sz_{j+1}$ type.
\section{Six-vertex model}
\label{sec:Six-vertex}
In this section we will find the quantum Hamiltonian in two approximations 
for the six-vertex model \cite{Sutherland_1970:jmp11-3183}.
The model consists of arrows on a two-dimensional square lattice.
Arrows are associated to each link between the nearest lattice sites.
The total energy is the sum over the lattice vertex energies.
Each vertex energy depends on configurations of four 
neighboring arrows. Moreover, only six of such configurations are allowed.
In Fig.1 the vertex configurations and the
appropriate energies are shown.

Not all possible arrow configurations are allowed: for each vertex the arrow 
configuration must be one of a set depicted in Fig.1 by bold solid arrows.
In this paper we will consider a lattice "rotated" by $\frac{\pi}{4}$ with 
periodic boundary conditions imposed. The total energy has the form
\begin{align}
E=\sum_{m=1}^M
&\sum_{n=1}^N\sidesum{odd}
E(\sigma_{m,n},\sigma_{m,n+1},\mu_{m,n},\mu_{m,n+1})
\nn
+&\sum_{n=1}^N\sidesum{even}
E(\mu_{m,n},\mu_{m,n+1},\sigma_{m+1,n},\sigma_{m+1,n+1}),
\end{align}
where the sum runs over all vertices.
Arrows directions are encoded by variables $\sigma$ and $\mu$.
We will denote the variables in the two nearest rows $m$ and $m+1$ by 
$\set{\sigma},\set{\mu}$ and 
$\set{\sigma'},\set{\mu'}$ (see Fig. 2).
The local energy for each vertex has the form
\begin{align}
\label{eqn:sixvertexElocal}
E(\sigma_1,\sigma_2,\mu_1,\mu_2)=
e_0 \overdelta_{\sigma_1}^{\sigma_2} \delta_{\sigma_1\mu_1} \delta_{\sigma_2\mu_2}
+e_x \delta_{\sigma_1\sigma_2} \delta_{\sigma_1\mu_1} \delta_{\sigma_2\mu_2}
+e_y \overdelta_{\sigma_1}^{\sigma_2} \overdelta_{\sigma_1}^{\mu_1}
\overdelta_{\sigma_2}^{\mu_2} \nn
+e_\infty \delta_{\sigma_1\sigma_2} \left(
 \delta_{\sigma_1\mu_1} \overdelta_{\sigma_2}^{\mu_2}
+\overdelta_{\sigma_1}^{\mu_1} \delta_{\sigma_2\mu_2}
+\overdelta_{\sigma_1}^{\mu_1} \overdelta_{\sigma_2}^{\mu_2}
\right)
+
e_\infty \overdelta_{\sigma_1}^{\sigma_2} \left(
 \delta_{\sigma_1\mu_1} \overdelta_{\sigma_2}^{\mu_2}
+ \overdelta_{\sigma_1}^{\mu_1} \delta_{\sigma_2}^{\mu_2}
\right),
\end{align}
where $\delta_{ab}$ is the Kronecker symbol,
$\overdelta_{a}^{b}=1-\delta_{ab}$.
All forbidden configurations appear in \eqref{eqn:sixvertexElocal} with 
energies $e_\infty$ which will be later sent to infinity and the
Boltzmann weights for the forbidden configurations will go to zero.
By using the local energy symmetry
$E(\mu_1,\mu_2,\sigma_1,\sigma_2)$ = $E(\sigma_1,\sigma_2,\mu_1,\mu_2)$
and the trivial identity $e_x\delta_{\sigma_1\sigma_2} \delta_{\sigma_1\mu_1}
\delta_{\sigma_2\mu_2}
\equiv
e_x \delta_{\sigma_1\sigma_2}
(1-\delta_{\sigma_1\mu_1} \overdelta_{\sigma_2}^{\mu_2}
-\overdelta_{\sigma_1}^{\mu_1} \delta_{\sigma_2\mu_2}
-\overdelta_{\sigma_1}^{\mu_1} \overdelta_{\sigma_2}^{\mu_2})$
we can divide energy in two parts
\begin{figure}
\centering
\psfrag{e_x}[tc][tc][1][0]{$e_x$}
\psfrag{e_y}[tc][tc][1][0]{$e_y$}
\psfrag{e_0}[tc][tc][1][0]{$e_0$}
\psfrag{e_i}[tc][tc][1][0]{$
{
\def\arraystretch{0.0}
\setlength\arraycolsep{0.0pt}
\begin{array}{cc}
e_\infty
\\
\overbrace{
\hspace*{0.6\textwidth}
}
\end{array}
}
$}
\includegraphics[width=\textwidth]{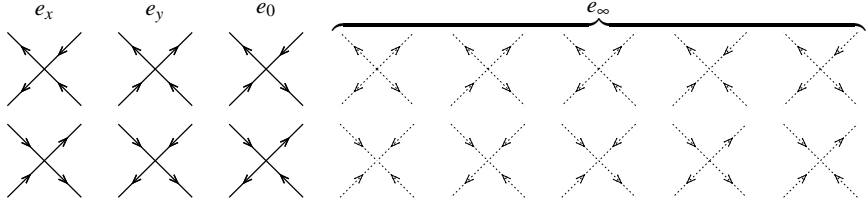}
\caption{All possible configurations of rows at a single vertex.
The six-vertex model admits only the first six configurations
(solid arrows) with finite energy.
All configurations prohibited for six-vertex model (dotted arrows)
have the energies that will be tend to infinity
and the Boltzmann weights become zero.}
\label{fig:sixvertexEnergies}
\end{figure}
\begin{align}
E_x(\sigma_1,\sigma_2,\mu_1,\mu_2) = e_x \delta_{\sigma_1, \sigma_2},
\nn
\label{eqn:sixvertexEy}
E_y(\sigma_1,\sigma_2,\mu_1,\mu_2) =
e_0 \overdelta_{\sigma_1}^{\sigma_2}\delta_{\sigma_1\mu_1}\delta_{\sigma_2\mu_2}
+ 0 \cdot \delta_{\sigma_1\sigma_2}\delta_{\sigma_1\mu_1}\delta_{\sigma_2\mu_2}
+ e_y\overdelta_{\sigma_1}^{\sigma_2}
\overdelta_{\sigma_1}^{\mu_1}\overdelta_{\sigma_2}^{\mu_2}
\nn
+ (e_\infty - e_x) \delta_{\sigma_1\sigma_2}
\left(
\delta_{\sigma_1\mu_1}\overdelta_{\sigma_2}^{\mu_2}
+\overdelta_{\sigma_1}^{\mu_1}\delta_{\sigma_2\mu_2}
+\overdelta_{\sigma_1}^{\mu_1}\overdelta_{\sigma_2}^{\mu_2}
\right)
\nn
+ e_\infty \overdelta_{\sigma_1}^{\sigma_2}
\left(
\delta_{\sigma_1\mu_1}\overdelta_{\sigma_2}^{\mu_2}
+\overdelta_{\sigma_1}^{\mu_1}\delta_{\sigma_2}^{\mu_2}
\right).
\end{align}
A reason for introducing the term proportional to zero in 
\eqref{eqn:sixvertexEy}
will be seen later.
Now, following our scheme we can write down the two components of 
the transfer matrix
\begin{align}
\label{eqn:sixvertexTx}
T^x_{\set{\sigma,\mu},\set{\sigma',\mu'}}
=
\prod_{n=1}^N
\delta_{\sigma_n \sigma_n'}
\delta_{\mu_n \mu_n'}
\emb{\frac{e_x}{2}\sum_{n=1}^N \delta_{\sigma_n \sigma_{n+1}}},
\\
\label{eqn:sixvertexTy}
T^y_{\set{\sigma,\mu},\set{\sigma',\mu'}}
=
\prod_{n=1}^N
\sidesum{odd}
\emb{E_y(\sigma_n,\sigma_{n+1},\mu_n,\mu_{n+1})}
\prod_{n=1}^N
\sidesum{even}
\emb{E_y(\sigma'_n,\sigma'_{n+1},\mu_n,\mu_{n+1})}.
\end{align}
Let us consider
$\emb{E_y(\sigma_{1},\sigma_{2},\mu_{1},\mu_{2})}$ in more details. 
As it can be seen in \eqref{eqn:sixvertexEy}, 
$E_y$ depends on four variables
$\sigma_{1},\sigma_{2},\mu_{1},\mu_{2}$. Each of these variables
can take two values $\pm\frac12$, and therefore, set of four variables
can take $2^4$ different sets of values.
For each set of values of $\sigma_1$, $\sigma_2$, $\mu_1$, $\mu_2$
one and only one term in \eqref{eqn:sixvertexEy} 
is nonzero, and therefore we can write
\begin{align}
\label{eqn:sixvertexEy1}
\emb{E_y(\sigma_{1},\sigma_{2},\mu_{1},\mu_{2})} =&
\emb{e_0} 
\overdelta_{\sigma_1}^{\sigma_2}\delta_{\sigma_1\mu_1}\delta_{\sigma_2\mu_2}
+\emb{0} \delta_{\sigma_1\sigma_2}\delta_{\sigma_1\mu_1}\delta_{\sigma_2\mu_2}
+\emb{e_y} \overdelta_{\sigma_1}^{\sigma_2}
\overdelta_{\sigma_1}^{\mu_1}\overdelta_{\sigma_2}^{\mu_2}
\nn
&+ \emb{e_\infty}
\e{\beta e_x} \delta_{\sigma_1\sigma_2}
(
\delta_{\sigma_1\mu_1}\overdelta_{\sigma_2}^{\mu_2}
+\overdelta_{\sigma_1}^{\mu_1}\delta_{\sigma_2\mu_2}
+\overdelta_{\sigma_1}^{\mu_1}\overdelta_{\sigma_2}^{\mu_2}
)
\nn
&+\emb{e_\infty}\overdelta_{\sigma_1}^{\sigma_2}
( \delta_{\sigma_1\mu_1}\overdelta_{\sigma_2}^{\mu_2}
+\overdelta_{\sigma_1}^{\mu_1}\delta_{\sigma_2\mu_2}
).
\end{align}
We are free to redefine the energies, 
$e_x \rightarrow e_x - e_0$, $e_y \rightarrow e_y - e_0$, 
$e_0 \rightarrow 0$, 
and to send the energies of forbidden configurations to infinity, 
$\emb{e_\infty} \rightarrow 0$. By denoting $t_y=\emb{e_y}$ we get
\begin{align}
\label{eqn:sixvertexEy2}
\emb{E_y(\sigma_{1},\sigma_{2},\mu_{1},\mu_{2})} 
= \delta_{\sigma_1\mu_1}\delta_{\sigma_2\mu_2}
+ t_y \overdelta_{\sigma_1}^{\sigma_2}\overdelta_{\sigma_1}^{\mu_1}
\overdelta_{\sigma_2}^{\mu_2}.
\end{align}
Now we have
\begin{align}
\label{eqn:sixvertexTy1}
T^y_{\set{\sigma,\mu},\set{\sigma',\mu'}} =
\prod_{n=1}^N\sidesum{odd} \left[
\delta_{\sigma_n\mu_n}
\delta_{\sigma_{n+1}\mu_{n+1}}
+ t_y \overdelta_{\sigma_n}^{\sigma_{n+1}}
\overdelta_{\sigma_n}^{\mu_n}\overdelta_{\sigma_{n+1}\mu_{n+1}}
\right]
\nn
\times
\prod_{n=1}^N\sidesum{even} \left[
\delta_{\sigma'_n\mu_n}
\delta_{\sigma'_{n+1}\mu_{n+1}}
+ t_y \overdelta_{\sigma'_n}^{\sigma'_{n+1}}
\overdelta_{\sigma'_n}^{\mu_n}\overdelta_{\sigma'_{n+1}}^{\mu_{n+1}}
\right]
\nn
=
\prod_{n=1}^N
\delta_{\sigma_n\mu_n}
\delta_{\sigma'_n,\mu_n}
+
t_y \sum_{j=1}^N \prod_{n=1}^N \nneqjjpo{n}{j}
\delta_{\sigma_n,\mu_n}
\delta_{\sigma'_n,\mu_n}
\nn
\times
\left\{
\begin{array}{ll}
\overdelta_{\sigma_j}^{\sigma_{j+1}}
\overdelta_{\sigma_j}^{\mu_j}
\overdelta_{\sigma_{j+1}}^{\mu_{j+1}}
\delta_{\sigma'_j\mu_j}
\delta_{\sigma'_{j+1}
\mu_{j+1}},
&\mbox{if $j$ odd}\\
\overdelta_{\sigma'_j}^{\sigma'_{j+1}}
\overdelta_{\sigma'_j}^{\mu_j}
\overdelta_{\sigma'_{j+1}}^{\mu_{j+1}}
\delta_{\sigma_j\mu_j}
\delta_{\sigma_{j+1}\mu_{j+1}},
&\mbox{if $j$ even}
\end{array}
\right.
+ \oty.
\end{align}
\begin{figure}
\centering
\psfrag{sigma_m_j}[cc][cl][0.75][-45]{$\sigma_{j}$}
\psfrag{sigma_m_jp1}[cc][cl][0.75][45]{$\sigma_{j+1}$}
\psfrag{mu_m_j}[cc][cl][0.75][45]{$\mu_{j}$}
\psfrag{mu_m_jp1}[cc][cl][0.75][-45]{$\mu_{j+1}$}
\psfrag{mu_m_jp2}[cc][cl][0.75][45]{$\mu_{j+2}$}
\psfrag{sigma_mp1_jp1}[cc][cl][0.75][45]{$\sigma'_{j+1}$}
\psfrag{sigma_mp1_jp2}[cc][cl][0.75][-45]{$\sigma'_{j+2}$}
\psfrag{sigma_mp1_jp3}[cc][cl][0.75][45]{$\sigma'_{j+3}$}
\psfrag{mu_mp1_jp2}[cc][cl][0.75][45]{$\mu'_{j+2}$}
\psfrag{mu_mp1_jp3}[cc][cl][0.75][-45]{$\mu'_{j+3}$}
\psfrag{sigma_m}[bl][bl][0.85][0]{$\set{\sigma_{m,j}}_{j=\overline{1,N}}\equiv \set{\sigma_j}_{j=\overline{1,N}}$}
\psfrag{mu_m}[bl][bl][0.85][0]{$\set{\mu_{m,j}}_{j=\overline{1,N}}\equiv \set{\mu_j}_{j=\overline{1,N}}$}
\psfrag{sigma_mp1}[bl][bl][0.85][0]{$\set{\sigma_{m+1,j}}_{j=\overline{1,N}}\equiv \set{\sigma'_j}_{j=\overline{1,N}}$}
\psfrag{mu_mp1}[bl][bl][0.85][0]{$\set{\mu_{m+1,j}}_{j=\overline{1,N}}\equiv \set{\mu'_j}_{j=\overline{1,N}}$}
\psfrag{rowm}[cl][cl][0.85][0]{$\left.\phantom{\rule[-0.7cm]{0.1cm}{1.0cm}}  \right\} \mbox{row } m$}
\psfrag{rowmp1}[cl][cl][0.85][0]{$\left.\phantom{\rule[-0.7cm]{0.1cm}{1.0cm}}  \right\} \mbox{row } m+1$}
\includegraphics[width=\textwidth]{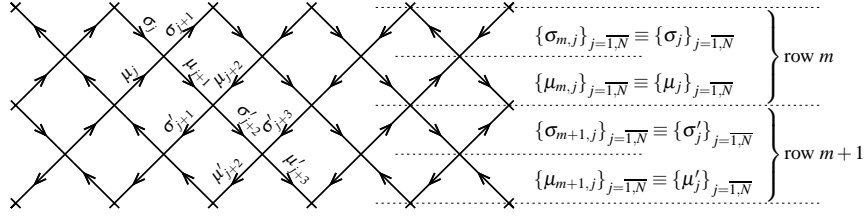}
\caption{Some permitted configurations for the six-vertex model.
We show two neighbor rows to explain a definition of variables
$\sigma$ and $\mu$.}
\label{fig:sixvertexRows}
\end{figure}
Using the identities $\delta_{ab}\delta_{bc}\equiv\delta_{ab}\delta_{ac},\;
\delta_{ab}\overdelta_{b}^{c}\equiv\delta_{ab}\overdelta_{a}^{c}$ we can rewrite 
$T^y$ in the form
\begin{align}
\label{eqn:sixvertexTy2}
T^y_{\set{\sigma,\mu},\set{\sigma',\mu'}} =
\prod_{n=1}^N
\delta_{\sigma_n,\sigma'_n}
\delta_{\sigma_n,\mu_n}
\nn
+
t_y \sum_{j=1}^N \prod_{n=1}^N \nneqjjpo{n}{j}
\delta_{\sigma_n\sigma'_n}
\delta_{\sigma_n\mu_n}
\overdelta_{\sigma_j}^{\sigma_{j+1}}
\overdelta_{\sigma_j}^{\sigma'_j}
\overdelta_{\sigma_{j+1}}^{\sigma'_{j+1}}
\times
\left\{
\begin{array}{ll}
\delta_{\sigma'_j\mu_j}
\delta_{\sigma'_{j+1}\mu_{j+1}},
&\mbox{if $j$ odd}\\
\delta_{\sigma_j\mu_j}
\delta_{\sigma_{j+1}\mu_{j+1}},
&\mbox{if $j$ even}
\end{array}
\right.
+ \oty.
\end{align}
The matrices $T$, $T^x$ and $T^y$ are of the size $2^{2N}\times2^{2N}$ and
the quantum Hamiltonian $\hH=-\tfrac{1}{\beta}\ln \hT$ should be defined in the
$2^{2N}$-dimensional Hilbert space. We can show that in the particular case
\eqref{eqn:sixvertexTy} the space dimension can be reduced to $2^{N}$. Let us 
recall that we are looking for thermodynamic properties of the system 
described by the partition function
$\Sp T^M$ \eqref{eqn:TpowerM}.
Matrix \eqref{eqn:sixvertexTy} can be represented in the form
$T^y = T^y_1 + t_y T^y_2 + \oty$ and by using the trace properties we may write
\begin{align}
\Sp T^M 
&=\Sp \left(T^x (T_1^y + t_y T_2^y + \oty) T^x\right)^M
\nn
&=\Sp \left[\left( (T^x)^2 T_1^y \right)^M 
+M t_y \left( (T^x)^2 T_1^y \right)^{M-1} (T^x)^2 T_2^y \right] + \oty.
\end{align}
The trace over space $\set{\sigma,\mu}$ can be splitted into two partial
traces over $\set{\sigma}$ and over $\set{\mu}$.
From Eqs. \eqref{eqn:sixvertexTx}, \eqref{eqn:sixvertexTy}
one can easy write down the matrix elements for
$\left((T^x)^2 T_1^y\right)^M$ and 
$M t_y \left((T^x)^2 T_1^y \right)^{M-1}T_2^y$
and perform the partial trace $\Sp_\set{\mu}$.
The result is as follows
\begin{align}
\label{eqn:sixvertexSpaceReduction}
\Sp T^M =& 
\Sp_\set{\sigma} \Sp_\set{\mu} \left[\left( (T^x)^2 T_1^y \right)^M + 
M t_y \left( (T^x)^2 T_1^y \right)^{M-1}(T^x)^2 T_2^y \right] + \oty
\nn
=&
\Sp_\set{\sigma}
\emb{M e_x \sum_j \delta_{\sigma_j\sigma_{j+1}}}
\left[
\prod_{n=1}^N \delta_{\sigma_n\sigma'_n}
+
M t_y \sum_{j=1}^N \prod_{n=1}^N \nneqjjpo{n}{j} \delta_{\sigma_n\sigma'_n}
\overdelta_{\sigma_j}^{\sigma_{j+1}}
\overdelta_{\sigma_j}^{\sigma'_j}
\overdelta_{\sigma_{j+1}}^{\sigma'_{j+1}}
\right]
\nn
&+\oty.
\end{align}
The latter expression in Eq. \eqref{eqn:sixvertexSpaceReduction}
is $\Sp_\set{\sigma} (\tilde{T}^x \tilde{T}^y \tilde{T}^x)^M+\oty$,
where the matrices $\tilde{T}^x$, $\tilde{T}^y$ are defined by their elements
\begin{align}
\label{eqn:sixvertexTildeTxTy}
\tilde{T}^x_{\set{\sigma},\set{\sigma'}}=&
\prod_{n=1}^N
\delta_{\sigma_n \sigma'_n}
\e{\mbox{-}\beta\frac{e_x}{2} \sum_{j=1}^N \delta_{\sigma_j\sigma_{j+1}}},
\nn
\tilde{T}^y_{\set{\sigma},\set{\sigma'}}=&
\prod_{n=1}^N\delta_{\sigma_n\sigma'_n}+ 
t_y \sum_{j=1}^N \prod_{n=1}^N \nneqjjpo{n}{j}
\delta_{\sigma_n\sigma'_n}
\overdelta_{\sigma_j}^{\sigma'_j}
\overdelta_{\sigma_j}^{\sigma_{j+1}}
\overdelta_{\sigma'_j}^{\sigma'_{j+1}}
+\oty.
\end{align}
We can look for the ground-state energy of Hamiltonian $\vec{\tilde{H}}$,
which acts in the $2^N$-dimensional Hilbert space $\set{\sigma}$ and is defined 
by the logarithm of the matrix $\tilde{T} = \tilde{T}^x \tilde{T}^y \tilde{T}^x$.
The reason that we can perform space dimension reduction is as follows: 
For any two states
$\set{\sigma}$, $\set{\sigma'}$, which differ by only one $\sigma_j$,
there is only one configuration $\set{\mu}$, for which the matrix element 
of $T$ remains non-vanishing in the limit $\oty=0$.
From here we will omit the tilde and use the notations $T$, $T^x$ and $T^y$ 
to denote the matrices that depend only on $\set{\sigma}$ variables.
Matrices $\tilde{T}^x$, $\tilde{T}^y$ 
\eqref{eqn:sixvertexTildeTxTy} are represented in the form suitable for
obtaining operators $\hX$ and $\hY$ mentioned in \eqref{eqn:TXY}.
\begin{align}
\label{eqn:sixvertexhxhyelements}
\hX =&
\sum_{j=1}^N
\sum_{\set{\sigma}}
\bigotimes_{n=1}^{N} \ket{\sigma_n}
\delta_{\sigma_j\sigma_{j+1}}
\bigotimes_{n=1}^{N} \bra{\sigma_n}
,
\nn
\hY =&
\sum_{j=1}^N
\sum_{\set{\sigma,\sigma'}}
\bigotimes_{n=1}^N \ket{\sigma_n}
\prod_{n=1}^N \nneqjjpo{n}{j}
\delta_{\sigma_n\sigma'_n}
\overdelta_{\sigma_j}^{\sigma'_j}
\overdelta_{\sigma_j}^{\sigma_{j+1}}
\overdelta_{\sigma'_j}^{\sigma'_{j+1}}
\bigotimes_{n=1}^N \bra{\sigma'_n}.
\end{align}
If the states defined by $\set{\sigma}$ are considered as the spin states
$\ket{\smallhalf}=\ket{\uparrow}$,
$\ket{\mbox{-}\smallhalf}=\ket{\downarrow}$, we can
immediately recognize the spin operators
\begin{align}
\label{eqn:sixvertexspinoperators}
\sum_{\sigma_j,\sigma_{j+1}} \ket{\sigma_j} \otimes  \ket{\sigma_{j+1}}
\delta_{\sigma_{j}\sigma_{j+1}}
\bra{\sigma_{j}} \otimes  \bra{\sigma_{j+1}}
=
2(\sz_j \sz_{j+1} + \nicefrac{1}{4})
\\
\sum_{\sigma_j,\sigma_{j+1}}
\sum_{\sigma'_j,\sigma'_{j+1}}
\ket{\sigma_j} \otimes  \ket{\sigma_{j+1}}
\overdelta_{\sigma_j}^{\sigma'_j}
\overdelta_{\sigma_j}^{\sigma_{j+1}}
\overdelta_{\sigma'_j}^{\sigma'_{j+1}}
\bra{\sigma'_j} \otimes  \bra{\sigma'_{j+1}}
=
2(\sx_j \sx_{j+1} + \sy_j \sy_{j+1}),
\end{align}
We have almost all that we need to construct the quantum Hamiltonian 
in both approximations.
At last we have only to recall Eq. \eqref{eqn:sixvertexTx} to define $t_x$
and collect essential variables and operators
\begin{align}
\label{eqn:sixvertextxtyhxhy}
t_x = -\frac{\beta e_x}{2},\;\;
t_y = \emb{e_y},\;\;
\hX = 2 \sum_{j=1}^N \sz_j \sz_{j+1} + \nicefrac{1}{4},\;\;
\hY = 2 \sum_{j=1}^N \sx_j \sx_{j+1} + \sy_j \sy_{j+1}.
\end{align}
In the strong-anisotropy limit \eqref{eqn:Hsal} we can 
immediately write the quantum Hamiltonian,
which corresponds to the spin-$\nicefrac{1}{2}$ $XXZ$ chain
\begin{align}
\label{eqn:sixvertexHsal}
\hH_{sal} = 2\sum_{j=1}^N \left( e_x \sz_j \sz_{j+1} -
\tfrac{1}{\beta}\emb{e_y}( \sx_j \sx_{j+1} + \sy_j \sy_{j+1})
\right),
\end{align}
where we omit the insignificant constants.
For the strong-interaction limit we have to calculate the commutator
\begin{align}
\label{eqn:sixvertexCommutator}
[\hX, \hY]_{2k} = 2^{2k} \sum_{j=1}^N
\left(1-4 \sz_{j-1} \sz_{j+2}\right)
(\sx_{j} \sx_{j+1} + \sy_{j} \sy_{j+1}).
\end{align}
This commutator has the form \eqref{eqn:Z_commutatorexplicitly} with
$P,Q=1$, $l=0$, $\vec{L}_{0,1} = \sum_j 
(1-4 \sz_{j-1} \sz_{j+2})$ $\cdot(\sx_{j} \sx_{j+1} + \sy_{j} \sy_{j+1})$,
$\vec{z}_{0,1} = 2$,
$\vec{R}_{0,1} = 1$.
Since $P=1$, we need calculate only one function in \eqref{eqn:apl} 
$\mathcal{A}_{0,1}(t_x) = \mathcal{A}(t_x)\equiv \mathcal{A}(-t_x)$, 
in order to construct operator $\hZ$ \eqref{eqn:ZviaA}. The Hamiltonian
\eqref{eqn:Hsil} becomes
\begin{align}
\label{eqn:sixvertexHsil}
\hH_{sil} = 2\sum_{j=1}^N e_x \sz_j \sz_{j+1} -
\tfrac{1}{\beta}\emb{e_y}
\left(
1+\frac{\mathcal{A}(\beta e_x)}{2}
\left(1-4 \sz_{j-1} \sz_{j+2}\right)
\right) (\sx_j \sx_{j+1} + \sy_j \sy_{j+1}).
\end{align}
\section{Conclusions}
\label{sec:Conclusions}
To compare the results provided by both approximations we will discuss the 
critical temperature for the Ising model which corresponds to quantum phase
transition of the Hamiltonians obtained in the strong-anisotropy limit 
\eqref{eqn:isingHsal} and in the strong-interaction limit \eqref{eqn:isingHsil}.
Both quantum \eqref{eqn:isingHsal} and \eqref{eqn:isingHsil} models 
are particular cases of generalized spin-$\smallhalf$ $XY$ chain 
for which a critical point can be found by Jordan-Wigner and Bogolubov 
transformations \cite{Suzuki_1971:ptp46-1337}.
Moreover, both quantum models \eqref{eqn:isingHsal} and \eqref{eqn:isingHsil}
have the critical point at the parameters which 
correspond to the same value for the critical temperature of the 
classical $2D$ Ising model
\begin{align}
\label{eqn:isingbetasil}
\frac{\beta_{sal}J_x}{4} \exp{\frac{\beta_{sal} J_y}{2}} = 1,\;\;
\beta_{sil}\equiv\beta_{sal}.
\end{align}

The fact that more accurate approximation, i.e. the strong-interaction 
approximation, does not give any improvement for
the value of the critical temperature can be explained based on analysis 
of the well-known exact equation for the Ising model critical temperature,
\begin{align}
\label{eqn:isingbetaex}
\sinh \frac{\beta_{ex}J_x}{2} \sinh \frac{\beta_{ex}J_y}{2} = 1.
\end{align}
Eq. \eqref{eqn:isingbetaex} can be rewritten in terms of $t_x$, $t_y$ 
\eqref{eqn:isingttxtyhxhy} as follows:
\begin{align}
\label{eqn:isingbetaexexpansion}
t_x = \mbox{arsinh} \left(\frac{2t_y}{1+t_y^2}\right) = 2t_y + 
{\mathcal O}(t_y^3).
\end{align}
In the strong-interaction limit $\oty=0$ and we have $t_x\propto t_y$
that is the requirement of the strong-anisotropy limit.
The fact that both gave the same critical temperature
can be explained in terms of
interactions $J_x$, $J_y$ and inverse temperature $\beta$.
In space of parameters $J_x$, $J_y$, $\beta$ 
the strong-interaction limit covers much wider region than 
the strong-anisotropy limit.
But near to the surface of critical temperatures defined 
by Eq. \eqref{eqn:isingbetaex} these regions coincide.
More precisely intersections of two surfaces given by exact solutions to
$\eqref{eqn:isingbetaex}$ and by the strong-interaction limit
$\emb{\frac{J_y}{2}}=0$ is a line defined by 
the strong-anisotropy limit $\emb{\frac{J_y}{2}}=0$, $\frac{\beta J_x}{2}=0$.
This transparent geometrical interpretation unfortunately can not be 
simply depicted because all regions which are object of our interest are 
infinitely distant.

For the six-vertex model the critical temperature in the strong-anisotropy limit 
can be calculated from isotropy condition for the $XXZ$ Hamiltonian,
\eqref{eqn:sixvertexHsal}
\begin{align}
\label{eqn:sixvertexbetasal}
\e{\mbox{-}\beta_{ex}e_y} = \beta_{ex}e_x.
\end{align}
The critical temperature in the strong-interaction limit requires
a study of the quantum spin chain described by Hamiltonian 
\eqref{eqn:sixvertexHsil} for which no exact results are available.
From the exact equation for the six-vertex model critical temperature,
\begin{align}
\label{eqn:sixvertexbetaex}
\e{\mbox{-}\beta_{ex}e_x}+\e{\mbox{-}\beta_{ex}e_y} = 1,
\end{align}
we can draw conclusions similar to those derived for the Ising model.
In fact, equation \eqref{eqn:sixvertexbetaex} can be
rewritten (see \eqref{eqn:sixvertextxtyhxhy}) as
\begin{align}
\label{eqn:sixvertexbetaexexpansion}
t_x = \frac{\ln (1-t_y)}{2} = -\frac{t_y}{2} + \oty,
\end{align}
wherefrom the equivalence of two approximations near the critical 
temperature may be expected.
In Fig. 3 the results for critical temperature given by 
strong-anisotropy approximation and 
strong-interaction approximation are shown.
We can see that the strong-anisotropy limit 
gives reasonable results even when the classical system 
become isotropic.
\begin{figure}
\centering
\psfrag{alpha}[Br][Br][1][0]{$\alpha$}
\psfrag{beta}[tr][cr][1][0]{$\bar{\beta}$}
\psfrag{Ising}[Bl][Bl][1][0]{\small Ising}
\psfrag{Sixvertex}[Bl][Bl][1][0]{\small Six-vertex}
\includegraphics[width=0.5\textwidth]{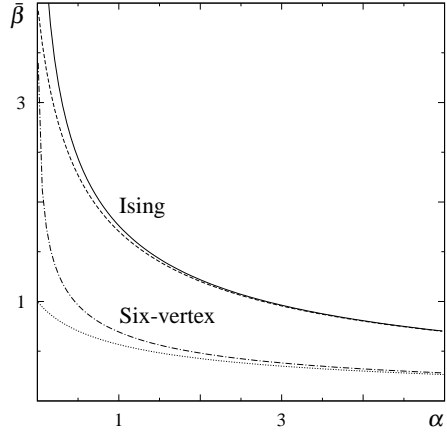}
\caption{Dependence of relative inverse critical temperature $\bar{\beta}$ 
on spatial anisotropy $\alpha$.
For the Ising model $\bar{\beta}=\beta J_x$, $\alpha=\frac{J_y}{J_x}$,
solid line corresponds to exact solution \eqref{eqn:isingbetaex},
dashed line corresponds to the strong-anisotropy and strong-interaction limits 
\eqref{eqn:isingbetasil}.
For the six-vertex model $\bar{\beta}=\beta e_x$, $\alpha=\frac{e_y}{e_x}$,
dash-doted line corresponds to exact solution \eqref{eqn:sixvertexbetaex},
doted line corresponds to the strong-anisotropy limit \eqref{eqn:sixvertexbetasal}.}
\label{fig:criticalTemperatures}
\end{figure}

In spite the fact that the novel strong-interaction approximation
does not improve the critical temperature of the considered classical
$2D$ models, the strong-interaction approximation have several advantages.
First, we can study isotropic classical systems. Second, we can apply 
strong-interaction approximation as low-temperature approximation 
for classical systems with arbitrary interactions.
Indeed we can achieve $t_y\propto\emb{e_y}$ 
to be small in two ways: by assuming strong interactions along one 
direction $e_y\rightarrow\infty$ or
putting low temperature $\beta\rightarrow\infty$.
\renewcommand{\theequation}{A\arabic{equation}}
\setcounter{equation}{0}
\appendix
\section*{Appendix: Representation of exponent products via nested commutators}
\label{sec:Appendix}
We will start from the expression for transfer matrix 
\eqref{eqn:Zintroducing},
where an unknown operator $\hZ$ was introduced
$\e{t_x \hX} (1+  t_y \hY + \oty) \e{t_x \hX} 
= \e{2t_x \hX+t_y (\hY+\hZ)} + \oty$.
By expanding the right-hand side
of the latter equation in powers of $t_y$ up to first order we derive
for the linear over $t_y$ terms the following result
\begin{align}
\e{t_x\hX} \hY \e{t_x\hX} =
\int_0^1 d \tau \e{2 \tau t_x \hX} (\hY+\hZ) \e{-2 \tau t_x \hX}
\e{2t_x \hX}.
\end{align}
The left and the right multiplication by $\e{-t_x \hX}$ and the substitution 
$\tau \to \frac{\tau + 1}{2}$,
$\int_0^1 \de \tau \to \tfrac{1}{2} \int_{-1}^1 \de \tau$
lead us to equation:
\begin{align}
\hY = \tfrac{1}{2} \int_{-1}^1 \e{\tau t_x \hX}
(\hY+\hZ) \e{-\tau t_x \hX}\de \tau  =
\tfrac{1}{2} \int_{-1}^1 \sum_{n=0}^\infty
\frac{\tau^n t_x^n \left( [\hX,\hY]_n + [\hX,\hZ]_n
\right)}{n!} \de \tau ,
\nn
{[\vec{A},\vec{B}]}_n = 
[\vec{A},{[\vec{A},\vec{B}]}_{n-1}],\;\;{[\vec{A},\vec{B}]}_0 = \vec{B}.
\end{align}
Now we can integrate over $\tau$
\begin{align}
\label{eqn:SILtxIdentities}
\hY = \tfrac{1}{2} \sum_{n=0}^\infty \frac{1+(-1)^n}{(n+1)!} t_x^n
\left( [\hX,\hY]_n + [\hX,\hZ]_n \right) =
\sum_{n=0}^\infty t_x^{2n} \frac{[\hX,\hY]_{2n} +
[\hX,\hZ]_{2n}}{(2n+1)!}.
\end{align}
Let us presume that operator $\hZ$ can be expanded 
in terms of nested commutators with unknown coefficients $A_k$:
$\hZ=\sum\limits_{k=1}^\infty \frac{A_k}{k!} t_x^k [\hX,\hY]_k$.
We can show that the function $Z(t_x\hX,\hY)$ is even with respect to 
the first argument,
\begin{align}
\begin{array}{rl}
\left.
\begin{array}{r}
\left( \e{t_x \hX} \e{t_y \hY} \e{t_x \hX} \right)^{-1}\!\!
=\!\left( \e{2t_x \hX+t_y (\hY+Z(t_x\hX,\hY))}\right)^{-1}\!\!
=\!\e{-2t_x \hX-t_y(\hY+Z(t_x\hX,\hY))}
\\
\left( \e{t_x \hX} \e{t_y \hY} \e{t_x \hX} \right)^{-1}\!\!
=\!\e{-t_x \hX} \e{-t_y \hY} \e{-t_x \hX}
=\!\e{-2t_x \hX-t_y(\hY+Z(-t_x\hX,\hY))}
\end{array}
\right\}
   & Z(t_x\hX,\hY)=Z(-t_x\hX,\hY),
\end{array}
\end{align}
that means that all odd $A_{2k+1}$ are zero $A_{2k+1}\equiv 0$.
Thus we will look for the following series for $\hZ$:
\begin{align}
\label{eqn:Z_viacommutators}
\hZ=\sum_{k=1}^\infty \frac{A_{2k}}{{2k}!} t_x^{2k}  [\hX,\hY]_{2k}
\end{align}
Substituting \eqref{eqn:Z_viacommutators} 
into \eqref{eqn:SILtxIdentities}
and reordering the sum (Cauchy product) 
$\sum\limits_{n=0}^\infty \sum\limits_{k=0}^\infty F(n,k)$ $=
\sum\limits_{i=0}^\infty \sum\limits_{j=0}^i F(j,i-j) $
we can write
\begin{align}
\hY
=
\sum_{i=0}^\infty \frac{1}{(2i+1)!} t_x^{2i} [\hX,\hY]_{2i}
+
\sum_{i=1}^\infty \left( \sum_{j=0}^{i-1} 
\frac{A_{2(i-j)}}{(2i - 2j)!(2j + 1)!} \right)
t_x^{2i}
[\hX,\hY]_{2i}.
\end{align}
Equating the coefficients at equal powers of $t_x$ we have the 
expressions for $A_{2k}$
\begin{align}
\sum_{j=0}^{i-1} \frac{A_{2(i-j)}}{(2i-2j)!(2j+1)!}+\frac{1}{(2i+1)!} = 0.
\end{align}
From this equations we can write a recursive representation for $A_k$,
\begin{align}
\label{eqn:Ak_recursivelly}
A_{2i}=-\frac{1}{2i+1} \sum_{j=1}^{i} \binom{2i+1}{2j+1} A_{2(i-j)},\;\;i 
\geq 1\;\;A_0=1,
\end{align}
where $\binom{2i+1}{2j+1} = \frac{(2i+1)!}{(2j+1)!(2i-2j)!}$.
It should be mentioned that for the Bernoulli numbers similar recursion 
representation exists,
\begin{align}
B_{2i}=-\frac{1}{2i+1} 
\sum_{j=1}^{i} \binom{2i+1}{2j+1} B_{2(i-j)}+\tfrac{1}{2},\;\;i 
\geq 1\;\;B_0=1, B_1=-\tfrac{1}{2}.
\end{align}

Our new task is to find the expansion \eqref{eqn:Z_viacommutators}.
Evidently it is impossible to do for arbitrary 
operators $\hX$ and $\hY$.
We will consider only the case in which the commutator
$[\hX,\hY]_{2k}$ have following periodical (with period $P$) structure:
\begin{align}
\label{eqn:Z_commutatorexplicitly}
[\hX,\hY]_{2Pr+2l} = \sum_{q=1}^Q \vec{L}_{l,q} 
\left(\vec{z}_{l,q}\right)^{2Pr+2l} \vec{R}_{l,q},
\end{align}
where $P,Q,r \in \mathbb{N}$, $l=\overline{0,P-1}$ and the
operators $\vec{L}_{l,q}\equiv\vec{L}_{l+P,q},\;
\vec{R}_{l,q}\equiv\vec{R}_{l+P,q},\;\vec{z}_{l,q}\equiv\vec{z}_{l+P,q}$.
Formula \eqref{eqn:Z_commutatorexplicitly} links nested commutators 
of the orders $2Pr+2l$, $r=\overline{1,\infty}$
by $Q$ rules. For each $l=\overline{0,P-1}$ this rules can be different.
With increasing the commutator order by $2P$ the expression for commutator 
is multiplied by operators $\vec{z}$ with 
constant sandwich multiplication by the operators
$\vec{L}$, $\vec{R}$.
If commutator $[\hX,\hY]_{2Pr+2l}$ have the form 
\eqref{eqn:Z_commutatorexplicitly} 
starting from $r'>1$, we can separate in 
\eqref{eqn:Z_viacommutators} the terms $k=\overline{1,r'-1}$ 
and perform all computations for the redefined operator $\hZ$.
Further progress is possible due to fact that an exponential generating 
function $\mathcal{A}(x)$ for the
coefficients $A_{2k}$ can be suggested
\begin{align}
\label{eqn:Z_A}
\mathcal{A}(x) \overset{\mbox{\tiny def}}{=} \sum_{k=1}^\infty \frac{A_{2k}}{2k!} x^{2k} 
\equiv \frac{x}{\sinh x} - 1.
\end{align}
From Eq. \eqref{eqn:Z_A} we can find a generating function for $\mathcal{A}_{P,l}(x)$,
\begin{align}
\label{eqn:apl}
\mathcal{A}_{P,l}(x)
\overset{\mbox{\tiny def}}{=}
\sum_{r=1}^\infty \frac{A_{2Pr+2l}}{(2Pr+2l)!} x^{2Pr+2l} =
\frac{1}{P} \sum_{p=0}^{P-1}
\e{-\frac{2i\pi p l}{P}}
\mathcal{A}\left(x \e{\frac{i \pi p}{P}}\right)
\end{align}
Finally, the unknown operator $\hZ$ takes the form
\begin{align}
\label{eqn:ZviaA}
\hZ = \sum_{q=1}^Q  \sum_{l=0}^{P-1} \vec{L}_{l,q}
\mathcal{A}_{P,l}(t_x \vec{z}_{l,q}) \vec{R}_{l,q}.
\end{align}
Having obtained this operator we can take logarithm of the transfer 
matrix \eqref{eqn:Zintroducing} and obtain quantum Hamiltonian 
in the strong-interaction limit \eqref{eqn:Hsil}

\end{document}